\documentclass{llncs}

\usepackage[toc,page]{appendix}
\usepackage{comment}
\usepackage{graphicx}
\graphicspath{{./Figs/}}
\usepackage{subfigure}
\usepackage{url}

\begin{document}
\title{An Adaptive Gas Cost Mechanism for Ethereum to Defend Against Under-Priced DoS Attacks}

\author{Ting Chen\inst{1,2} \and Xiaoqi Li\inst{2} \and Ying Wang\inst{1} \and Jiachi Chen\inst{2} \and Zihao Li\inst{1} \and Xiapu Luo\inst{2}\thanks{The corresponding author} \and Man Ho Au\inst{2} \and Xiaosong Zhang\inst{1}}
\institute{Cyber Security Research Center,
	University of Electronic Science and Technology of China,
	Chengdu 611731, China\\
	\email{\{brokendragon|johnsonzxs\}@uestc.edu.cn, \{769836805|gforiq\}@qq.com}
	\and
	Department of Computing,
	Hong Kong Polytechnic University,
	Hong Kong,\\
	\email{\{csxqli|chenjiachi317\}@gmail.com, \{csxluo|csallen\}@comp.polyu.edu.hk}
    }

\maketitle

\begin{abstract}
The gas mechanism in Ethereum charges the execution of every operation to ensure that smart contracts running in EVM (Ethereum Virtual Machine) will be eventually terminated. Failing to properly set the gas costs of EVM operations allows attackers to launch DoS attacks on Ethereum. Although Ethereum recently adjusted the gas costs of EVM operations to defend against known DoS attacks, it remains unknown whether the new setting is proper and how to configure it to defend against unknown DoS attacks. In this paper, we make the \emph{first} step to address this challenging issue by first proposing an emulation-based framework to automatically measure the resource consumptions of EVM operations. The results reveal that Ethereum's new setting is still not proper. Moreover, we obtain an insight that there may always exist exploitable under-priced operations if the cost is fixed. Hence, we propose a novel gas cost mechanism, which dynamically adjusts the costs of EVM operations according to the number of executions, to thwart DoS attacks. This method punishes the operations that are executed much more frequently than before and lead to high gas costs. To make our solution flexible and secure and avoid frequent update of Ethereum client, we design a special smart contract that collaborates with the updated EVM for dynamic parameter adjustment.  Experimental results demonstrate that our method can effectively thwart both known and unknown DoS attacks with flexible parameter settings. Moreover, our method only introduces negligible additional gas consumption for benign users. 
\end{abstract}

\section{Introduction}
\label{sec_intro}

Being the second largest blockchain~\cite{ether_second}, Ethereum distinguishes itself by its Turing-complete execution environment(i.e, EVM)~\cite{white_paper} that can run various applications through smart contracts. Besides transferring money, transactions in Ethereum are also involved in deploying and invoking smart contracts. To ensure that the execution of smart contracts will be terminated eventually, Ethereum charges \emph{gas} (i.e., execution fee) from transaction senders, and lets it be part of the rewards to miners for executing smart contracts. In particular, gas serves as a protection mechanism against resources abusing in case executing certain smart contracts consumes lots of computing resources. The money paid for executing an EVM operation (e.g., addition, multiplication, reading the balance of an account) is the multiplication of the \emph{gas price} with the \emph{gas cost} of that operation, where the gas price indicates the value of one unit of gas and the gas cost of an EVM operation stands for the units of gas required to execute the operation. The gas cost is determined by the EVM in Ethereum client, and the gas price can be set by transaction senders. Every transaction has a \emph{gas limit}, dubbed TGL (Transaction Gas Limit), so that the execution of a smart contract will trigger an \emph{out-of-gas} exception if the execution requires more gas than the TGL. Ethereum attempts to associate EVM operations' gas costs proportionally to the computing resources needed to execute them\cite{white_paper}, because a proper setting of gas costs can give miners proper awards and thwart DoS attackers who aim at wasting a large amount of resources. 

However, it is non-trivial to properly set the gas cost of each operation because it requires a deep understanding of EVM internals, an accurate measurement of resource consumptions by EVM operations, and the awareness of the market price for different types of computing resources (e.g., CPU, memory, etc.). Failing to select suitable gas costs for EVM operations gives attackers opportunities to launch DoS attacks on Ethereum at low cost by exploiting under-priced operations. An operation is regarded as under-priced if its gas cost is lower than what it should be. Actually, two DoS attacks exploiting such operations were discovered in 2016, which repeatedly execute two under-priced operations, namely \texttt{EXTCODESIZE}~\cite{ext_attack} and \texttt{SUICIDE}~\cite{suicide_attack}, thus resulting in slow transaction processing, wasted hard drive space, and long synchronization time. More seriously, the confidence of users in Ethereum will be shaken, and consequently the market price of Ethereum will be impacted~\cite{market}. Since each Ethereum node should maintain the complete copy of blockchain and replay all transactions in history for synchronization, such DoS attacks happened in history will also impact the newly enrolled nodes. Although Ethereum adjusted the gas costs of operations to defend against such known attacks~\cite{ext_attack,suicide_attack}, it remains unknown whether or not the new setting is resistant to unknown attacks and how to properly configure the gas costs of operations to mitigate DoS attacks.

In this paper, we make the \emph{first} step to address this challenging issue by first proposing an emulation-based framework (in Section \ref{sec_measure}) to automatically measure the consumptions of computing resources of EVM operations. The framework consists of the interpretation handler for each EVM operation, the related data structures and diverse simulated environments in an attempt to explore all program paths of those handlers. The experimental result reveals that the latest setting in Ethereum is still not proper although it can mitigate the known DoS attacks. From this investigation, we obtain the insight that there may always exist exploitable under-priced operations if the operation costs are fixed, because the factors influencing the costs of EVM operations keep changing.

Therefore, we propose a novel adaptive gas cost mechanism (in Section \ref{sec_price}), which will dynamically adjust the costs of EVM operations according to the number of executions, to defend against known and unknown DoS attacks. This mechanism punishes the operations leading to abnormal high gas costs if they are executed much more frequently than before. Consequently, the exponentially increased gas costs will impede the attackers without unlimited money from conducting effective DoS attacks. Our experiments in a private blockchain show that the new mechanism can effectively thwart both known and unknown DoS attacks and introduce negligible additional gas consumption to benign users.

Moreover, by exploiting Ethereum's unique feature, we realize our mechanism through a novel approach in order to make it secure and flexible in terms of parameter adjustment. More precisely, we develop a specific smart contract and provide a patch to EVM. After patching the EVM, the developers of Ethereum can adjust the parameters by sending transactions to that smart contract, and then the updated EVM can fetch the parameters periodically by reading the storage of that smart contract. Our new approach leverages the underlying blockchain technique to make the parameters auditable and untamperable. Moreover, our approach has good deployability because it only needs updating the EVM once.

In summary, we make the following major contributions:

\noindent(1) We propose the first emulation-based measurement framework, which can automatically estimate the resource consumptions of EVM operations, to assess whether or not the gas costs in Ethereum are properly configured (Section \ref{sec_measure}). 

\noindent(2) We propose a novel adaptive gas cost mechanism, which dynamically adjusts operation costs according to their execution times, to defend against known and unknown DoS attacks with negligible impacts on benign users.

\noindent(3) We design a new approach to realize our gas cost mechanism by exploiting Ethereum's smart contract and its underlying bloackchain technique. This approach makes our mechanism secure, flexible, easy to be deployed.

\noindent(4) We conduct experiments in a private blockchain to evaluate our mechanism. The results show that it can effectively thwart both known and unknown DoS attacks and introduce negligible additional gas consumption to benign users. Moreover, the parameters can be dynamically adjusted by authorized users.

The remainder of this paper is organized as follows. Section \ref{sec_background} introduces background knowledge. Section \ref{sec_analysis} presents our analysis of two real DoS attacks on Ethereum. Section \ref{sec_measure} details the measurement framework. We describe the adaptive gas cost mechanism and its implementation in Section \ref{sec_price} and Section \ref{sec_implement}, respectively. The experiment results are introduced in Section \ref{sec_evaluation}. After summarizing related studies in Section \ref{sec_related}, we conclude the paper in Section \ref{sec_conclusion}.

\section{Background}
\label{sec_background}
This section introduces some background knowledge of Ethereum. Besides providing a cryptocurrency (i.e., Ether), Ethereum supports deploying and running smart contracts. There are two types of accounts in Ethereum, including external owned accounts (EOA) and smart contracts. The major difference between them is that only smart contracts contain executable bytecode~\cite{document}. Ethereum uses the underlying P2P overlay to deliver transactions among Ethereum nodes. A \emph{transaction} refers  to the signed data package that stores a message to be sent from an EOA to another account on the blockchain~\cite{document}. A block is a data structure to store zero or more transactions. Each node runs an Ethereum client that obeys Ethereum protocol~\cite{yellow_paper}. The consensus in Ethereum is achieved by using a modified version of GHOST protocol~\cite{white_paper}, and as the result of the consensus, every node maintains the same copy of the blockchain. In particular, a newly joined node should download all blocks (i.e., synchronization) and then run all historical transactions to reach the same state as the other nodes.

Ethereum can be considered as a state machine where a state is a snapshot of the blockchain (e.g., the balances of all accounts, the value of a variable in a smart contract) and a transaction results in a state transfer. If the target of a transaction is a smart contract, the smart contract will be executed in EVM. Since EVM is usually embedded in the Ethereum client, the execution of smart contracts consumes the computing resources (e.g., CPU, disk, network) of each node. Consequently, a DoS attack will impact all nodes because each of them should execute all historical transactions. To prevent abusing computing resources, Ethereum leverages gas to charge execution fee from transaction senders. The amount of gas consumption is determined by the executed EVM operations, and different operations may have different gas costs~\cite{yellow_paper}. In Section \ref{sec_analysis}, we use real attacks to explain how attackers exploit the improper setting of gas cost to launch DoS attacks at low expense.

\section{Analyzing Real DoS Attacks on Ethereum}
\label{sec_analysis}
This section dissects two real DoS attacks exploiting under-priced operations. 

\subsection{\texttt{EXTCODESIZE} Attack}
\label{sec_ext}
\noindent\textbf{Approach:} The attacker sends lots of transactions to invoke a deployed smart contract involving many \texttt{EXTCODESIZE} operations, which gets the size of an account's code~\cite{yellow_paper}. Such attack forces \texttt{EXTCODESIZE} to be executed roughly 50,000 times per block~\cite{ext_attack}. 

\noindent\textbf{Symptom:} Clients spend a very long time to process those blocks that contain the transactions sent from the attacker, and hence the throughput of Ethereum for processing transactions is decreased.

\noindent\textbf{Cause:} \texttt{EXTCODESIZE} has a very low gas cost (i.e., 20 in go-ethereum V 1.3.5), but it involves expensive operation (i.e., reading information from the disk). Hence, the execution of a great number of \texttt{EXTCODESIZE} results in busy I/O and slow transaction processing speed.

\noindent\textbf{Countermeasure:} New Ethereum (e.g., go-ethereum V 1.6) increases the gas cost of \texttt{EXTCODESIZE} to 700~\cite{eip150} (in the source file gas\_table.go). 
Consequently, the  transaction senders have to pay 35 ($=700/20$) times more money when using go-ethereum V 1.6. 700 gas is equal to about 0.000014 Ether (many senders set the gas price to 0.00000002 Ether at August, 2017), whose value is about 0.0042 USD (1 Ether can be exchanged into about 300 USD at August 13th, 2017~\cite{usd}). Although a single operation does not cost much, the accumulative gas consumption is considerable, because each transaction incurs the execution of many operations and there are more than 45 million transactions from the launch of Ethereum to August 13th, 2017~\cite{etherscan}.

\subsection{\texttt{SUICIDE} Attack}
\label{sec_suicide}
\noindent\textbf{Approach:} The attacker creates lots of smart contracts with a loop in their constructors. In the loop, the \texttt{SUICIDE} operation is executed. According to Ethereum's protocol, \texttt{SUICIDE} is used to remove the executed smart contract from the blockchain and send the remaining Ether to the designated account~\cite{yellow_paper}. For each generated smart contract, the transaction for creating it triggers its constructor, and hence lots of \texttt{SUICIDE} whose target accounts do not exist, will be executed. Note that a nonexistent account does not need to be stored in the Ethereum state tree~\cite{suicide_attack}, which represents the state of the blockchain.

\noindent\textbf{Symptom:} About 19 million accounts were created by the attack, which consume considerable disk space, and thus the synchronization and transaction processing are slowed down.

\noindent\textbf{Cause:} If the target account does not exist, a \texttt{SUICIDE} operation will turn it into existent, which will be stored in the Ethereum state tree~\cite{suicide_attack}. However, the gas cost of \texttt{SUICIDE} is zero. Therefore, an attacker creates a huge number of accounts by executing \texttt{SUICIDE} repeatedly at very low cost.

\noindent\textbf{Countermeasure:} New Ethereum increases the gas cost of \texttt{SUICIDE} to 5,000 and additional 25,000 if it creates a new account~\cite{eip150} (in the source file gas\_table.go). Moreover, new clients can delete the zombie accounts created by the attack.

\subsection{Remarks}
\label{sec_observation}
From the above analysis, we learn that to exploit the under-priced operations for launching DoS attacks, the attacker has to first find or prepare a smart contract containing the under-priced operations, and then cause such operations to be executed lots of times by sending transactions to the smart contract. Moreover, since the gas cost for sending a transaction is high (e.g., at least 21,000 in go-ethereum V 1.6), the attacker usually lets each transaction trigger multiple executions of the under-priced operations. To defend against such DoS attacks, we should either properly set the costs of EVM operations (i.e., remove under-priced operations) or force the attacker to pay a lot of money for executing the under-priced operations many times.

In Section \ref{sec_measure}, we propose a novel emulation-based measurement framework to assess whether or not the latest gas cost setting is proper. Unfortunately, we find that the latest setting still has exploitable under-priced operations, and it is difficult, if not impossible, to eliminate all under-priced operations if the operation costs are fixed, because the factors influencing the cost of each EVM operation keep changing. Therefore, we explore an alternative approach by proposing a novel adaptive gas cost mechanism in Section \ref{sec_price}.

\section{Emulation-based Measurement Framework}
\label{sec_measure}
Although Ethereum has changed the gas costs of some under-priced operations to defend against the known DoS attacks~\cite{ext_attack,suicide_attack}, little is known whether or not the latest gas cost setting is immune to DoS attacks exploiting under-priced operations. To address this issue, the resource consumption of each EVM operation should be measured. However, it is non-trivial to measure the computing resources consumed by a single EVM operation because the execution of a smart contract involves not only many EVM operations but also various utility functions for supporting the execution.

To tackle this problem, we propose a novel emulation-based measurement framework. More precisely, by exploiting EVM's architecture, we extract the interpretation handler for each operation (e.g., the \emph{opAdd()} function is responsible for executing the addition operation), the related data structures (e.g., stack, memory, storage) from the EVM implementation, and prepare an emulated environment, which consists of the \emph{Go} compiler, runtime libraries (e.g., the \emph{bigInt} library to handle large integers) and the state of the blockchain (e.g., the balance of an account), for executing the operation. Then, we run the interpretation handler in the emulated environment millions of times, because a single run is too short to conduct the measurement, and record the execution time. Note that the current implementation of our framework can automatically measure the CPU consumption in terms of the execution time, and we will support the measurement of other resources in future work.

\begin{figure}[ht]
	\centering
	\vspace{-5ex}
	\includegraphics[width=0.7\textwidth]{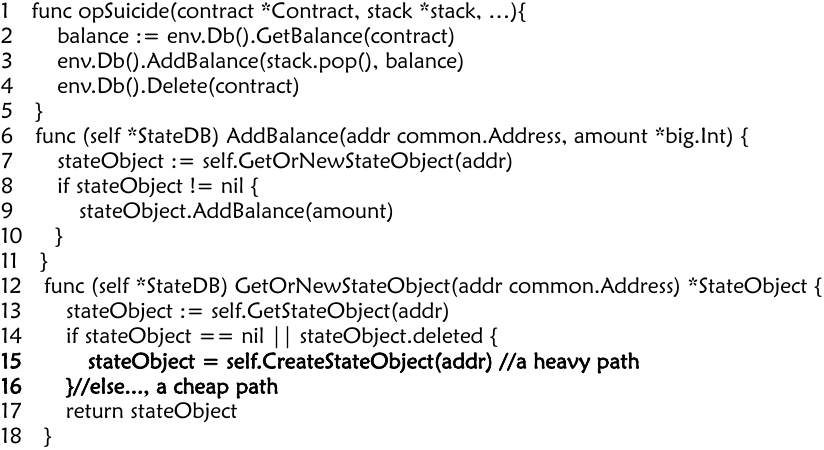}
	\vspace{-2ex}
	\caption{An expensive path and a cheap path of \emph{opSuicide}}
	\vspace{-5ex}
	\label{fig_path}
\end{figure}

There is a challenge in preparing the emulated environment. In particular, since a handler may have various execution paths with different resource consumption, we need to explore all execution paths for measuring the handler's resource consumption. The example in Fig .\ref{fig_path} shows that the handler for executing \texttt{SUICIDE} consists of an expensive path (Line 15, \emph{CreateStateObject()} allocates disk space to store accounts) if the target account is nonexistent since it will become existent after executing \texttt{SUICIDE} and a cheap path (Line 16) if the target is existent. To address this challenge, we run the handler millions of times, providing with different inputs and proper runtime environment. If the operation manipulates the stack/memory/storage, we synthesize the stack/memory/storage with random length and generates random numbers as their items. If the operation needs the information from EVM (e.g., block number, gas price, gas limit) or the smart contract  (e.g., code length, input to the contract, the address of the contract), we prepare an EVM/smart contract object with randomly generated fields. If the operation needs to interact with another account, we take into account the following three situations. First, if the target account is nonexistent, no special preparation is needed. Second, if the target account is an EOA, we generate one using the command provided by Etheruem's client. Third, if the target account is a smart contract, we develop and deploy one in the private chain, whose code is a \texttt{RETURN} since we measure the resources consumed by the invocation, rather than the execution of the invoked smart contract.

We classify all EVM operations into five categories in terms of the data structures on which they operate. The operations in the first category do not manipulate any data structures (e.g., \texttt{JUMPDEST}). The operations in the second category handle the stack (e.g., \texttt{ADD}). The operations in the third category get access to the specific fields related to blockchain (e.g., \texttt{ORIGIN}). The fourth category of operations manipulates the memory (e.g., \texttt{MSTORE}). The operations in the fifth category manipulate the storage (e.g., \texttt{SLOAD}).  Note that in Ethereum \emph{memory} is an infinitely expandable byte-array that resets after computation ends, while \emph{storage} is a long-term key/value store that persists for the long term~\cite{white_paper}.

\begin{figure}[ht]
	\centering
	\vspace{-2ex}
	\includegraphics[width=0.7\textwidth]{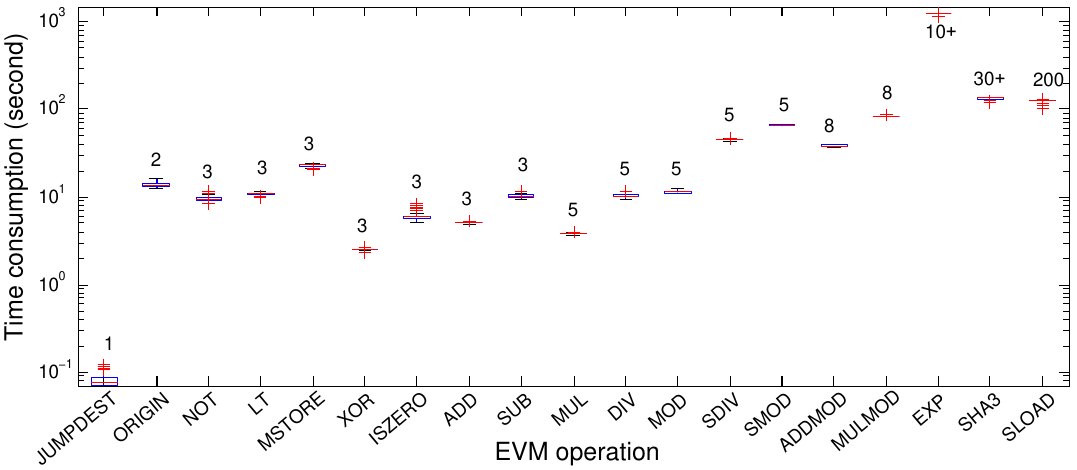}
	\vspace{-2ex}
	\caption{Time consumptions of EVM operations}
	\vspace{-3ex}
	\label{fig_measure}
\end{figure}

Fig. \ref{fig_measure} (the $y$-axis is on a log scale) presents the CPU consumptions of some EVM operations running 50 million times from all the five categories. Experiments are conducted on a desktop equipped with an Intel i3-4160 CPU and 8GB memory. The number on top of each box is the operation's gas cost according to Ethereum's yellow paper~\cite{yellow_paper}. All measurements repeat 100 times. \texttt{JUMPDEST} is the destination of a jump (e.g., \texttt{JUMP}, \texttt{JUMPI}) operation, which belongs to the first category. \texttt{ADD}, \texttt{SUB}, \texttt{MUL}, \texttt{DIV}, \texttt{SDIV}, \texttt{MOD}, \texttt{SMOD}, \texttt{ADDMOD}, \texttt{MULMOD} are arithmetic operations. \texttt{NOT} and \texttt{XOR} are bitwise operations. \texttt{ISZERO} and \texttt{LT} are comparison operations. These operations belong the second category. \texttt{ORIGIN} is the representative of the third category which reads a field of the block's head. \texttt{MSTORE} and \texttt{SHA3} belong to the fourth category. \texttt{MSTORE} writes a word to memory while \texttt{SHA3} can operate multiple items in memory. In particular, \texttt{SHA3} hashes the data in memory and its gas cost is the summation of basic gas (i.e., 30) with the gas for operating memory. The more memory it reads, the more gas it requires. \texttt{EXP} is a special arithmetic operation whose gas cost is the summation of basic gas (i.e., 10) with the remaining part which is determined by the bit length of the exponent. In other words, the gas cost of \texttt{EXP} becomes high if it has a large exponent. Fig. \ref{fig_measure} shows that \texttt{EXP} costs considerable CPU resources. \texttt{SLOAD} loads an item from the storage.

\begin{figure}[ht]
	\centering
	\vspace{-4ex}
	\includegraphics[width=0.8\textwidth]{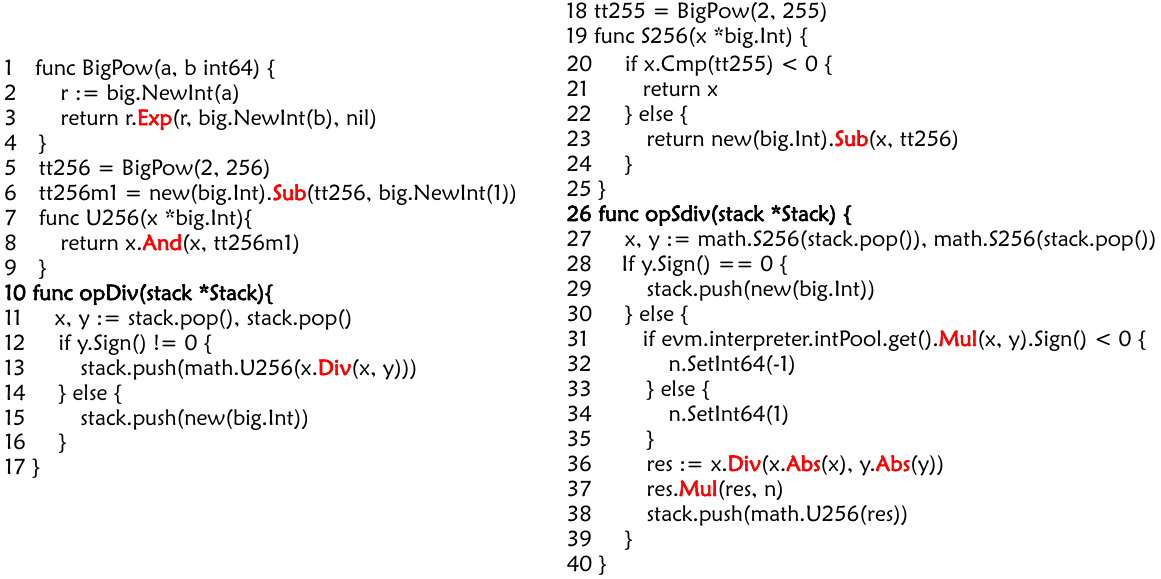}
	\vspace{-1ex}
	\caption{EVM source code for executing \texttt{DIV} and \texttt{SDIV}}
	\vspace{-4ex}
	\label{fig_div_sdiv}
\end{figure}

The results show that the latest gas costs are not proportional to the consumptions of CPU resources. For example, \texttt{DIV} (division) has the same gas cost of \texttt{SDIV} (signed division), but the execution time needed by \texttt{DIV} is about 23\% ($10.4s/45s$) of that needed by \texttt{SDIV}. We find the reason by investigating the source code of handlers for \texttt{DIV} and \texttt{SDIV}, which is listed in Fig. \ref{fig_div_sdiv}. Fig. \ref{fig_div_sdiv} lists the source code (from go-ethereum V 1.6) of division (function \emph{opDiv()}, Line 10) and signed division (function \emph{opSdiv()}, Line 26), respectively, whose gas costs are equivalent. For the  ease of presentation, we simplify the source code. We can see that the functions \emph{opDiv()} and \emph{opSdiv()} consist of stack operations (e.g., \emph{stack.pop()}) and math computations (e.g. \emph{x.And()}) provided by the \emph{bigInt} library. Further experiments reveal that math computations (in red color) take up most of the execution time.
We also find that the execution of a division operation needs 4 math computations (i.e., 1 \emph{Div}, 1 \emph{And}, 1 \emph{Sub} and 1 \emph{Exp}) at most whereas the execution of a signed division needs 11 (i.e., 3 \emph{Sub}, 3 \emph{Exp}, 2 \emph{Abs}, 1 \emph{Mul}, 1 \emph{Div} and 1 \emph{And}) at most. Hence, \texttt{SDIV} is more resource-consuming than \texttt{DIV}. Consequently, some operations (e.g., \texttt{EXP}, \texttt{SHA3}, as shown in Fig. \ref{fig_measure}) may be under-priced and thus could be exploited by DoS attacks.

\section{Adaptive Gas Cost Mechanism}
\label{sec_price}

The investigation in Section \ref{sec_measure} shows that it is not easy to properly assign gas costs to EVM operations. Hence, we propose a novel adaptive gas cost mechanism for defending against DoS attacks.

\subsection{Threat Model}
\label{sec_model}

We assume that the attacker can discover under-priced operations (if any) and then launch the attack by invoking either existing smart contracts or new smart contracts crafted by the attacker. Moreover, the attacker is rational and does not have unlimited money for launching attacks. In this case, she will give up the attack if her money cannot force the under-priced operations to be executed for lots of times. Moreover, she will not send a transaction that can execute the under-priced operations only a few times because sending a transaction is not cheap (i.e., gas cost is 21,000). Last but not least, normal users will not accept a gas cost mechanism that charges much money from them.

\subsection{Adaptive Adjustment of Gas Costs}
\label{sec_method}
Exploiting the observation in Section \ref{sec_observation} that a successful DoS attack has to trigger lots of executions of under-priced operations, we propose a new mechanism that increases the gas cost of an operation dynamically if it has been executed much more frequently than before.
More precisely, we collect the execution traces (i.e., a sequence of executed operations) of normal transactions, and model the execution frequency of each EVM operation. Then, for every new transaction, we set a basic gas cost for each operation by default, and count the number of executions of each operation. If the number of an operation is larger than a threshold, its gas cost will be increased. The advantage of our mechanism is that it does \emph{not} need to know which operations are under-priced. Instead, it punishes the over-frequent EVM operation through the increased gas cost. Hence, it can defend against known and unknown DoS attacks.

We define a threshold $\mu_i$ for the operation $i$ as shown in Eqn. (\ref{eq_u}). The operation $i$ that has been executed for more than $\mu_i$ in \emph{one} transaction is regarded as over-frequent, and its gas cost will be increased. $ave_i$ and $std_i$ stand for the average and the standard deviation of the number of executing operation $i$, respectively. Section \ref{sec_computation} details how to compute them. $base\_count$ is an integer used to prevent increasing the gas cost of an infrequently-executed operation too fast.  $m$ is a parameter for adjusting the threshold.

\vspace{-2ex}
\begin{equation}\label{eq_u}
\small
\mu_i=max\{base\_count, ave_i+m\times std_i\}
\end{equation}
\vspace{-2ex}

\begin{figure}[ht]
	\centering
	\vspace{-4ex}
	\includegraphics[width=0.4\textwidth]{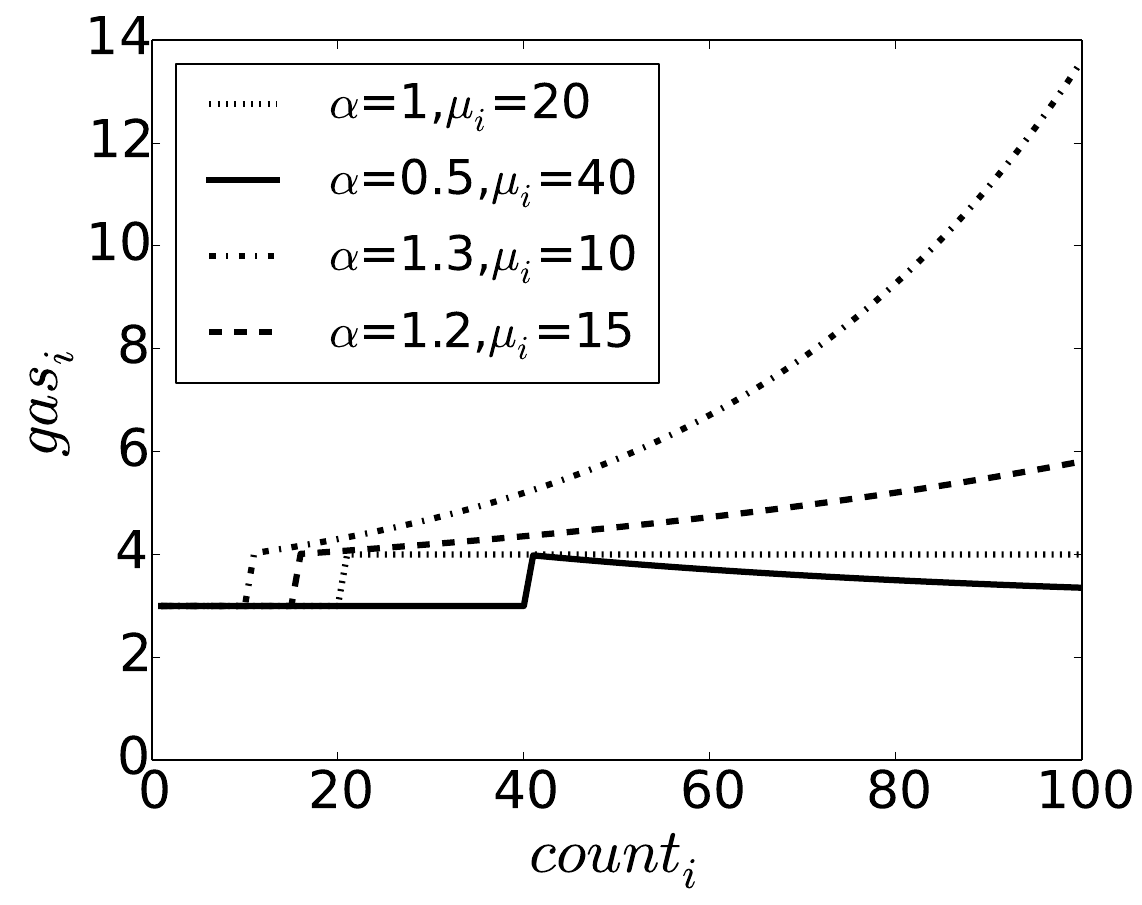}
	\vspace{-3ex}
	\caption{Curves of Eqn. (\ref{eq_gas}), $base\_gas_i=3$}
	\vspace{-4ex}
	\label{fig_exponent}
\end{figure}

The gas cost of an EVM operation is dynamically adjusted according to Eqn. (\ref{eq_gas}), where $count_i$ is the number of executions of operation $i$, $base\_gas_i$ is the default gas cost of $i$. We uses an exponential function to punish over-frequent operations with accelerating increments in gas costs. Its base (i.e. $\alpha > 1$) determines the speed of increasing the gas cost. We let the exponent as $\frac{count_{i}}{\mu_i}-1$ that includes $\mu_i$ for taking into account the operation's normal frequency. Since our mechanism will assign an operation a very high gas cost if it has been executed much more times in a transaction than before, it deters an attacker from executing an under-priced operation many times by one transaction. Moreover, our mechanism avoids charging much more gas from benign senders by setting proper parameters. We evaluate the effects of various parameters in Section \ref{sec_set}. $gas_i$ is restored to $base\_gas_i$ for a new transaction, and hence the attacker cannot affect the initial operation costs of benign transactions. Fig. \ref{fig_exponent} shows the curves of Eqn. (\ref{eq_gas}) with various parameters, indicating that $\mu_i$ and $\alpha$ can affect the point from where to increase gas cost and the speed to increase gas cost, respectively. We have several observations. First, $\mu_i$ determines the point from where $gas_i$ should be increased. Moreover, $\alpha$ determines the increasing speed of $gas_i$. Typically, $gas_i$ should be increased with the increase of execution number $count_i$, and hence $\alpha$ should be larger than 1.

\vspace{-2ex}
\begin{equation}\label{eq_gas}
\small
gas_i=\left\{
\begin{array}{l}
base\_gas_i,\ \ \ \ \ \ \ \ \ \ \ \ \ \ \ \ \ if\ count_{i} \leq \mu_i \\
base\_gas_i+\alpha ^{\frac{count_{i}}{\mu_i}-1},\ if\ count_{i} > \mu_i\\
\end{array}
\right.
\end{equation}
\vspace{-1ex}

Section \ref{sec_adaptive} will describe the way to adjust the parameters in Eqn. (\ref{eq_u}) and Eqn. (\ref{eq_gas}), and we will try other functions (e.g., linear, polynomial) in Eqn. (\ref{eq_gas}) in future work.

\subsection{Dynamic Parameter Configuration}
\label{sec_adaptive}

Since Ethereum and its smart contracts evolve over time, the parameters should be changed accordingly. Therefore, we need an approach for dynamic parameter configuration. This approach should meet the following requirements. First, the parameter configuration should be auditable by any users of Ethereum. Second, the parameter configuration should be secure so that attackers cannot modify the parameters. Third, the approach should not need to frequently update Ethereum client due to the risk of hard fork.

Exploiting Ethereum's unique feature, we propose a novel approach for realizing dynamic parameter configuration by developing a specific smart contract and providing a patch to EVM. The developers of Ethereum can adjust the parameters by sending transactions carrying new parameters to that smart contract. They can adjust a variable, \emph{block number}, in the smart contract, which is used to determine when the new setting takes effect. Then, the patched EVM can fetch the parameters periodically by reading the storage of that smart contract. The period (measured by blocks) of querying new parameters should be shorter than the difference between the variable, block number, in the smart contract and the block number when setting the new parameters so that all clients can get the newest setting before the block when the setting takes effect.

Our new approach leverages the underlying blockchain technique to make the parameters auditable and untamperable. Note that no one can change the setting of gas costs by just subverting her EVM. Moreover, the smart contract for updating parameters cannot be tampered by attackers who do not have more than 50\% computing power because the contract itself will be validated in the process of consensus. The change of parameters will be auditable because all transactions are publicly available in the blockchain. Last but not least the Ethereum client (i.e., its EVM) should only be updated once for adopting our new gas cost mechanism. After that, they do not need to be updated again for using the new parameters.
\section{Implementation}
\label{sec_implement}
\begin{figure}[ht]
	\centering
	\vspace{-6ex}
	\includegraphics[width=0.6\textwidth]{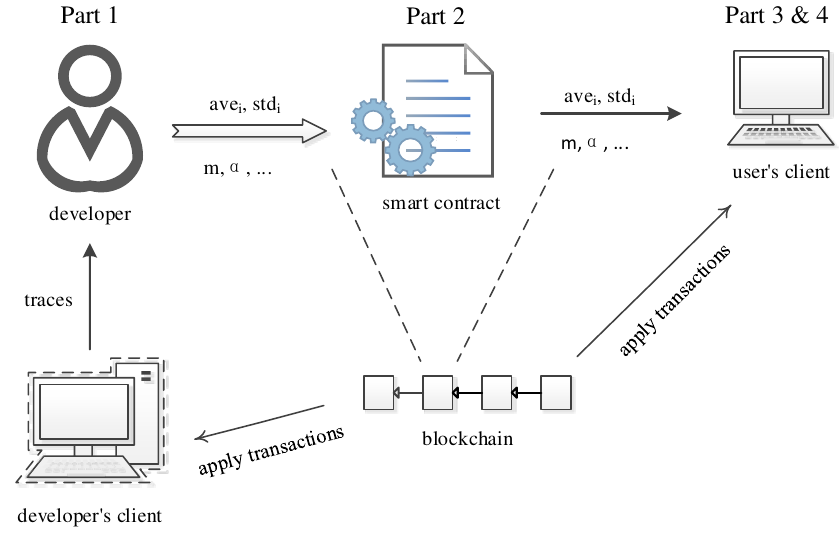}
	\vspace{-2ex}
	\caption{Overview of our implementation.}
	\vspace{-4ex}
	\label{fig_archi}
\end{figure}

The implementation of our new mechanism consists of four parts (Fig. \ref{fig_archi}). The first part collects execution traces of smart contracts and computes $ave_i$ and $std_i$. Part 2 is the smart contract storing the parameters that can be updated by Ethereum developers. Part 3 and 4 describe the patch to EVM, including how to fetch new parameters and how to apply them, respectively. 

\subsection{Computing $ave_i$ and $std_i$}
\label{sec_computation}
\begin{figure}[h!]
	\centering
	\vspace{-6ex}
	\subfigure[\texttt{Ave}]{
		\includegraphics[width=0.46\textwidth]{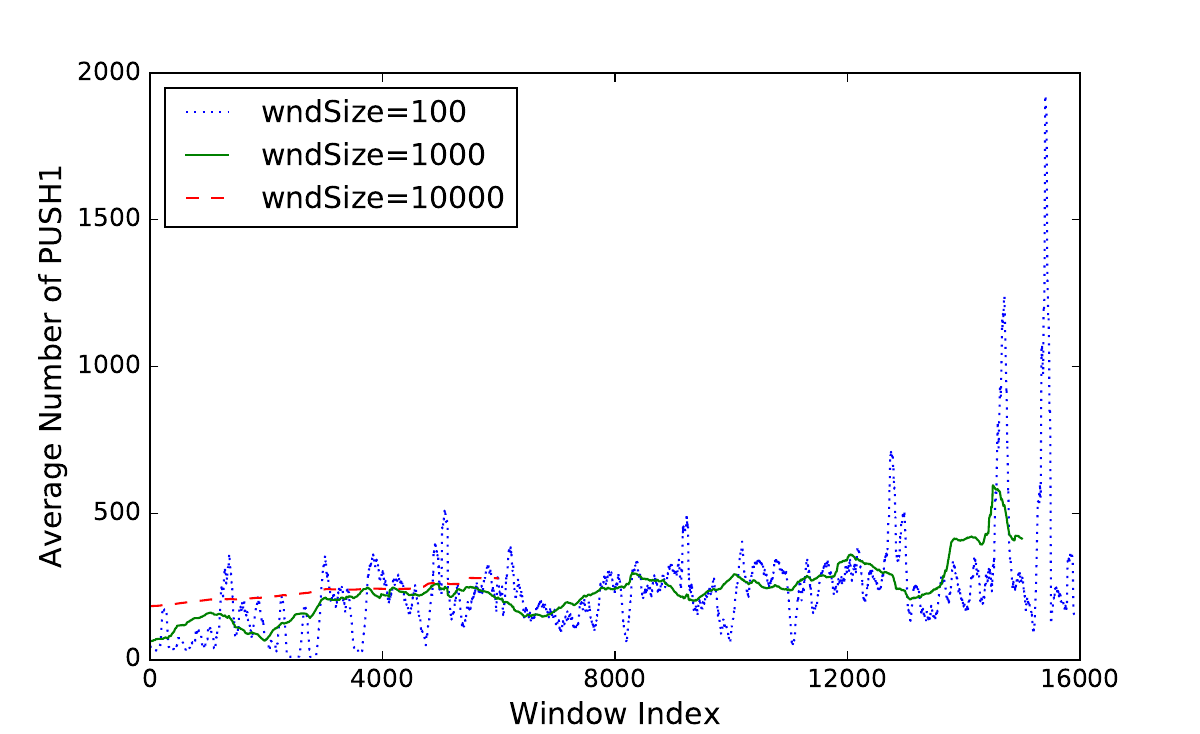}
	}	
	\subfigure[\texttt{Std}]{
		\includegraphics[width=0.46\textwidth]{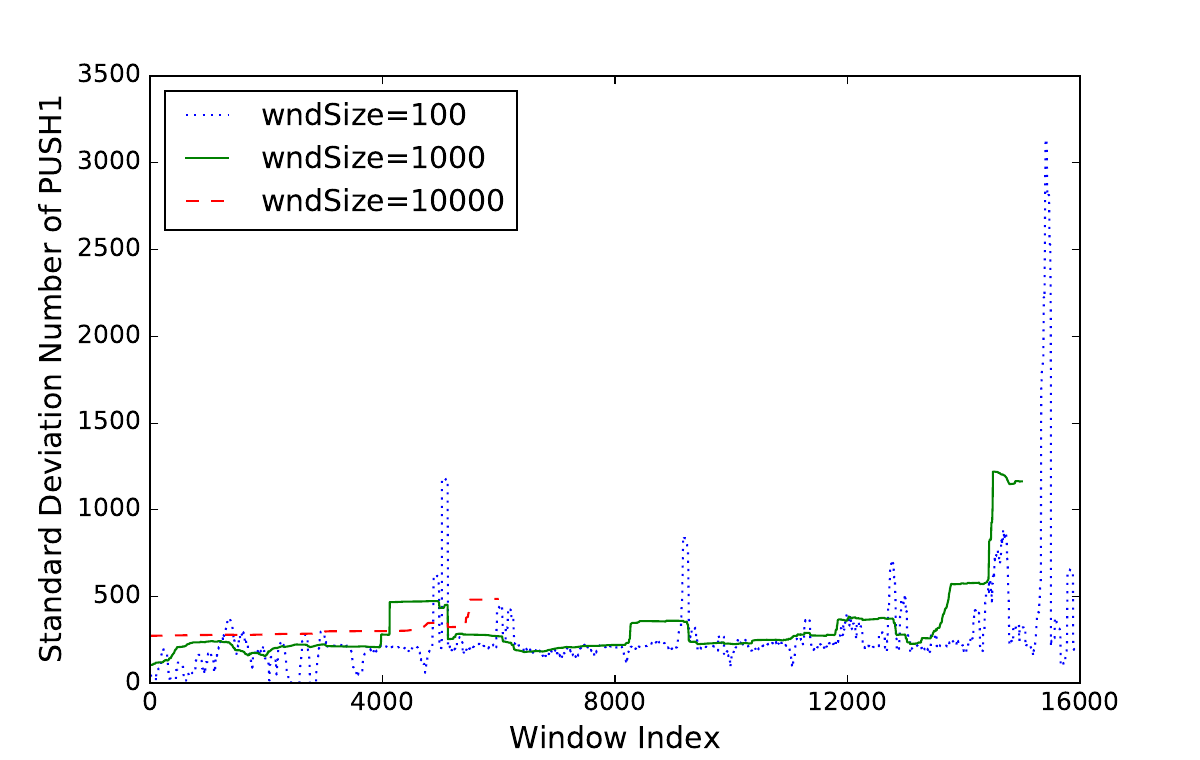}
	}	
	\vspace{-3ex}
	\caption{Average number and standard deviation of the executions of \texttt{PUSH1}.}
	\vspace{-4ex}
	\label{fig_ave_std}
\end{figure}

To compute $ave_i$ and $std_i$, we first leverage EVM's built-in tracing ability to record all execution traces. We define a sliding window, and use all traces within that window for computing $ave_i$ and $std_i$. Fig \ref{fig_ave_std} shows $ave_i$ and $std_i$ of \texttt{PUSH1} with different window sizes (i.e., 100, 1,000 and 10,000) in the first 16,000 execution traces since the launch of Ethereum. We assume that these traces were triggered by benign transactions since no known attacks were discovered in them. Please note that \texttt{PUSH1} is the most frequent operation, which pushes one byte on stack. 

The $x$-axis gives the window index and for example, a point ($x$, $y$) on the red line of Fig \ref{fig_ave_std}(a) indicates that $ave_i$ of \texttt{PUSH1} of the traces within the window [$x+1$, $x+10,000$] is $y$. We can see that the $ave_i$ of \texttt{PUSH1} increases as time goes on, indicating that smart contracts become more complicated than before. Second, as we expected, the larger the window is, the more stable $ave_i$ and $std_i$ will be. Moreover, it is difficult for an attacker to tamper $ave_i$ and $std_i$ by filling the large window with crafted transactions. Our approach allows developers to adjust the window size.

\subsection{Smart contract}
\begin{figure}[ht]
	\centering
	\vspace{-4ex}
	\includegraphics[width=0.6\textwidth]{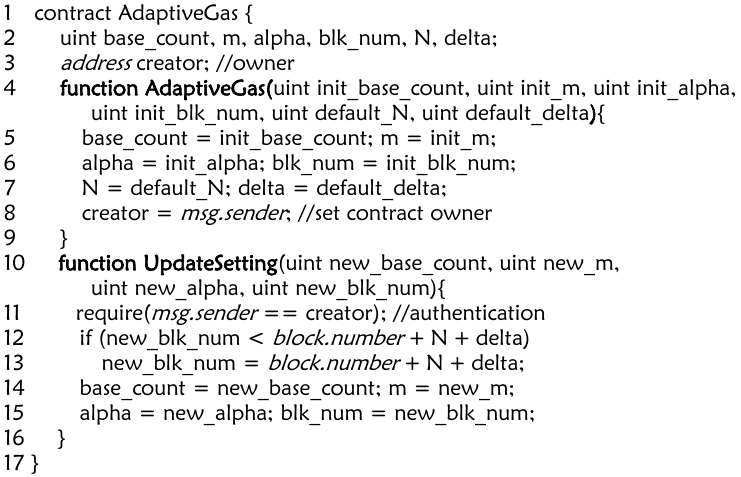}
	\vspace{-2ex}
	\caption{The smart contract for updating the setting of parameters}
	\vspace{-4ex}
	\label{fig_contract}
\end{figure}

We implement a smart contract (as shown in Fig. \ref{fig_contract}) to store parameters which allows the contract's creator to update parameters through executing transactions, and then we deploy it on our private blockchain. For ease of presentation, we omit the details of updating $ave_i$ and $std_i$ of each operation $i$, which is the same as the updating the other parameters (e.g., $m$). Line 2 declares several global variables which store in the storage. \emph{address} (Line 3) is a built-in variable type of Ethereum which can only be used for storing account address. $N$ is the time interval of two consecutive queries, and $delta$ is a small number that we consider all clients can get the new setting in the time period of $N+delta$ (Line 13). The function \emph{AdaptiveGas()} is the construct function that will be executed during the creation of the smart contract. Please note that the arguments of \emph{AdaptiveGas()} are also given in the transaction for contract creation.

Besides setting the default parameters in \emph{AdaptiveGas()} (Lines 5-7), we record the contract owner (Line 8), ensuring that only the owner can change parameters setting (Line 11). The function \emph{UpdateSetting()} accepts the new setting of parameters from transaction senders. Lines 12, 13 ensure that the time period ($N+delta$) is enough for all clients to check the update. Please note that \emph{msg.sender} and \emph{block.number} are two built-in properties of Ethereum that get the address of transaction sender and the number of block which contains the transaction, respectively. Please note that the transaction fees for sending the transactions to adjust paramters are negligible for Ethereum official society because a single transaction does not cost much (always less than 1 USD~\cite{etherscan}) and parameters do not need to adjust very frequently.

\subsection{Querying new parameters}
\begin{figure}[ht]
	\centering
	\vspace{-4ex}
	\includegraphics[width=0.4\textwidth]{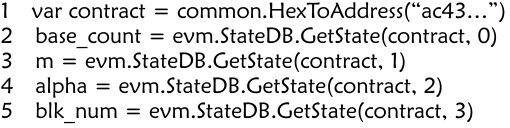}
	\vspace{-3ex}
	\caption{Modifications of EVM to obtain new parameters.}
	\vspace{-8ex}
	\label{fig_query}
\end{figure}

Fig. \ref{fig_query} shows the code snippet (simplified for presentation) for an EVM to get the new setting of parameters. Since each Ethereum node keeps a complete copy of blockchain, their EVMs can get the values of all storage variables given the address of the smart contract by accessing the local copy of blockchain. It is more efficient than an intuitive approach that fetches the new parameters by sending a transaction to the smart contract, because the latter will add transactions to the blockchain periodically and cause additional fee for sending transactions. Our approach can avoid these issues. Line 1 specifies the address (i.e., \emph{ac43...}) of the contract, which is known because the contract is developed and deployed by us. Then, Lines 2-5 obtain individual parameters by directly accessing (i.e., invoking the internal function \emph{evm.StateDB.GetState()} of EVM) the storage of the contract. The integers 0, 1, etc. give the locations of parameters stored in the storage. Finally, those parameters are used for computing gas costs.

\subsection{Applying new parameters}
We modify go-ethereum V 1.3.5 to realize our mechanism because it has several known under-priced operations, and we compare the original V 1.3.5 with the patched one in Section \ref{sec_eval_dos}. When Ethereum starts, we load the setting of parameters (e.g., $ave_i$, $std_i$, $m$, $\alpha$) in the entry function (i.e., the \emph{main()} function in \verb|\|cmd\verb|\|geth\verb|\|main.go). Please note that the default gas cost of each operation (i.e., $base\_gas_i$) is the same as that in go-ethereum V 1.3.5. We replace the code in the function \emph{CalculateGasAndSize()} in \verb|\|core\verb|\|vm\verb|\|vm.go, which is responsible for computing the gas consumption of individual operation, with our code to calculate gas cost and increase the execution number of the EVM operation by one. In other words, Eqn. (\ref{eq_u}) and Eqn. (\ref{eq_gas}) are implemented in \emph{CalculateGasAndSize()}. The number of executions will be reset before the execution of each transaction, which is implemented in the function \emph{ApplyTransaction()} in \verb|\|core\verb|\|state\_processor.go. To reduce the runtime overhead, we cache the gas costs of operations, which have already been computed, in main memory.

\section{Evaluation}
\label{sec_evaluation}
This section answers the following questions through experiments.

\noindent\textbf{RQ1:} Can our mechanism thwart known and unknown DoS attacks effectively?

\noindent\textbf{RQ2:} How much additional gas will be charged from benign users by our mechanism? 

\noindent\textbf{RQ3:} What are the effects of parameters? 

All experiments are conducted in a private Ethereum blockchain on a desktop equipped with an Intel Xeon E312 processor and 8GB memory. Our private blockchain has one miner and is isolated with the public Ethereum blockchain and other testing blockchains. We create an account to hold the rewards from mining. We guarantee that the account has enough money to send transactions by setting a low mining difficulty. Every block also has a gas limit, dubbed BGL (Block Gas Limit), which restricts the size of a block (i.e., the number of transactions contained in the block). The BGL is set as 4 million, which is comparable with that in the public chain at present. We let the TGL be equal to the BGL, in order to see how many under-priced operations can be executed by a single transaction using the original gas cost mechanism and our mechanism, respectively. The parameters $base\_count$, $m$ and $\alpha$ in Eqn. (\ref{eq_u}) and Eqn. (\ref{eq_gas}) are set to 5, 3 and 2 by default, respectively. We evaluate our mechanism under different settings in Section \ref{sec_set}.

\subsection{Experiments with DoS Attacks}
\label{sec_eval_dos}

We simulate the two real attacks~\cite{ext_attack,suicide_attack} in our private blockchain. To launch the \texttt{EXTCODESIZE} attack, we develop a smart contract with a public function \emph{extAttack()} that can be called by our account. \emph{extAttack()} has a loop where we use inline assembly to execute \texttt{EXTCODESIZE} directly. The \texttt{SUICIDE} attack is launched in a more intricate way since a smart contract will be removed (i.e., cannot get accessed) by executing \texttt{SUICIDE}. The \texttt{SUICIDE} attack exploits the feature of Ethereum: a smart contract will not be removed before the completion of the transaction that triggers the \texttt{SUICIDE} operation. Consequently, we create a smart contract whose constructor invokes \texttt{SUICIDE} in a loop. When creating the contract, the corresponding transaction executes \texttt{SUICIDE} repeatedly. We also use the built-in tracing ability of EVM to record the execution traces of smart contracts as well as the gas consumption of each operation.

\begin{figure}[ht]
	\centering
	\vspace{-4ex}
	\includegraphics[width=0.7\textwidth]{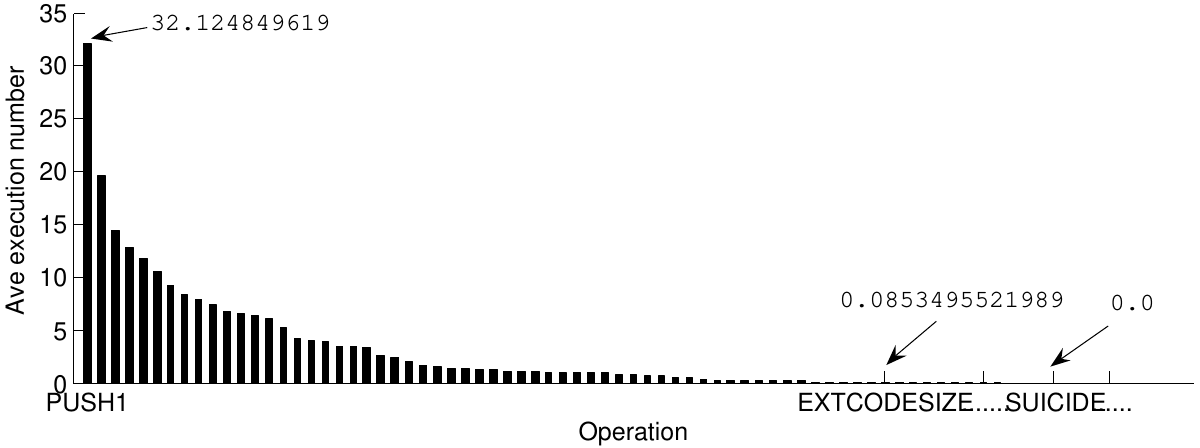}
	\vspace{-2ex}
	\caption{Average execution number of every EVM operation, 10,000 benign transactions collected from 07:40:00 AM, April 28, 2017 to 01:58:56 PM, April 28, 2017}
	\vspace{-4ex}
	\label{fig_ave}
\end{figure}

\begin{figure}[h!]
	\centering
	\vspace{-4ex}
	\subfigure[\texttt{EXTCODESIZE}]{
		\includegraphics[width=0.46\textwidth]{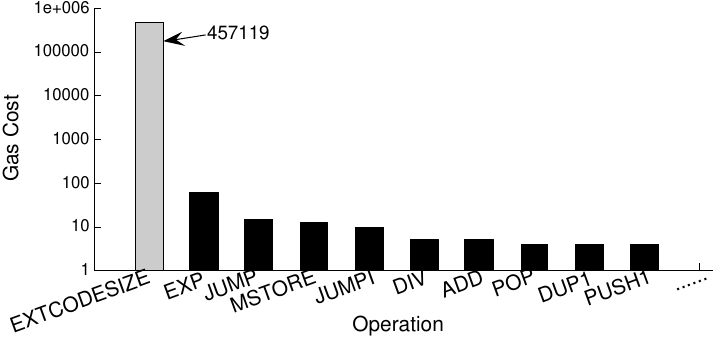}
	}	
	\subfigure[\texttt{SUICIDE}]{
		\includegraphics[width=0.46\textwidth]{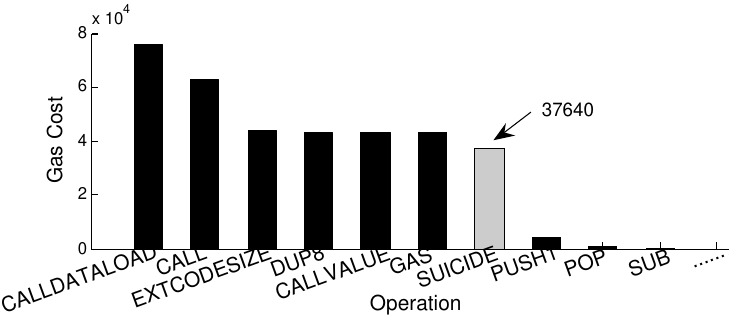}
	}	
	\vspace{-2ex}
	\caption{Gas of each operation after attacking}
	\vspace{-5ex}
	\label{fig_gas_attack}
\end{figure}

The experimental results reveal that the two attacks execute 92,494 and 11,335 times of \texttt{EXTCODESIZE} and \texttt{SUICIDE}, respectively, in one transaction using the original (i.e., go-ethereum V 1.3.5) gas cost mechanism. By contrast, the two attacks only execute 99 and 81 times of \texttt{EXTCODESIZE} and \texttt{SUICIDE}, respectively, with the same cost (i.e., 4 million gas) after our mechanism is applied. Fig. \ref{fig_gas_attack}(a) ($y$-axis is on a log scale) and Fig. \ref{fig_gas_attack}(b) shows the gas cost of each operation in descending order after the attacks when our mechanism used. Note that these two figures do not include all operations due to the page limit. We can see that the gas costs of the two under-priced operations become very expensive (i.e., 457,119 and 37,640 respectively) after attacks. We also find some other expensive operations (e.g., \texttt{CALLDATALOAD}, \texttt{CALL}) because they are also in the loop, resulting in over-frequent executions than before. Fig. \ref{fig_ave} demonstrates that the execution frequencies of different operations vary. Moreover, the two under-priced operations (i.e., \texttt{EXTCODESIZE} and \texttt{SUCIDE}) exploited by real attacks are rarely executed by benign users.

\begin{figure}[ht]
	\centering
	\vspace{-4ex}
	\includegraphics[width=0.5\textwidth]{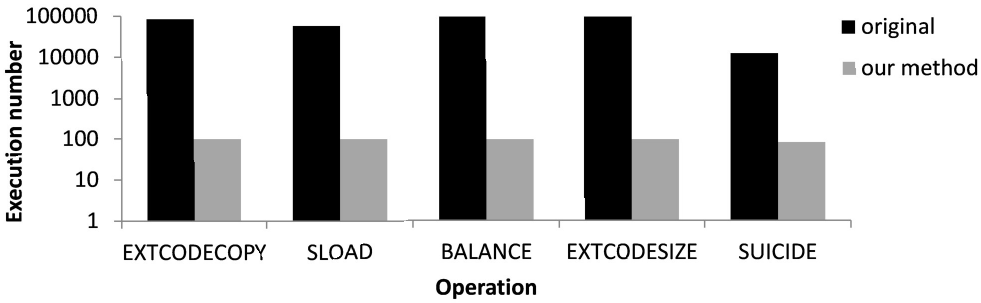}
	\vspace{-3ex}
	\caption{Execution numbers of under-priced operations}
	\vspace{-5ex}
	\label{fig_number}
\end{figure}

To evaluate our approach against unknown DoS attacks, we synthesize three attacks by executing three under-priced operations (i.e., \texttt{EXTCODECOPY}, \texttt{SLOAD} and \texttt{BALANCE}) in a loop, respectively, which are similar to the \texttt{EXTCODESIZE} attack. Note that go-ethereum V 1.3.5 will be affected by the DoS attacks exploiting these operations whereas the latest version of Ethereum has increased their gas costs. Fig. \ref{fig_number} demonstrates that our method reduces the number of executions of under-priced operations by several orders of magnitude. Therefore, the answer to RQ1 is:

\noindent
\fbox{%	
	\parbox{0.99\textwidth}{%
		\emph{Our gas cost mechanism can effectively thwart known and unknown DoS attacks.}
	}%	
}

\subsection{Experiments with Normal Transactions}
\label{sec_eval_normal}
To evaluate how much additional gas will be charged from normal users by our mechanism, we first randomly select 10 smart contracts and then replay their transactions in the original go-ethereum V 1.3.5 and the updated go-ethereum V 1.3.5 with our gas cost mechanism, respectively. 84 transactions in total are replayed, and 15 transactions (2 to one contract and the other 13 to another contract) out of them incur additional gas by our mechanism. The total gas consumed by 84 transactions under the original gas cost setting is 2,441,340, and the total additional gas incurred by our mechanism is 444. Therefore, the percentage of additional gas charged from benign users is about 0.018\%.

As a case study, we detail the experiment with one smart contract. More precisely, the smart contract is deployed at 0x61F9d1cE56aC1623FeD4e949D7D4202\\51fef0896. We compile the source and deploy the smart contract in the private blockchain. There are 37 transactions to that smart contract in total until April 29, 2017. 
We do not replay the transaction for contract creation since it does not trigger the execution of any public functions provided by the smart contract, nor the 4 transactions with internal transactions because our private blockchain is isolated from other accounts. Note that an internal transaction is not a real transaction and will not be stored in the blockchain. Instead, it is made by calling (via \texttt{CALL}, \texttt{CALLCODE}, \texttt{DELEGATECALL} etc.) an account from a smart contract. We also skip the transaction running out of gas, and hence we replay 31 ($37-1-4-1$) transactions.

The results show that 18 out of 31 transactions consume the same amount of gas under our mechanism and the original mechanism. The total increment in gas consumption of the other 13 transactions incurred by our mechanism is 130, and the largest increment in gas consumption of one transaction is 10. Please note that the total gas consumption of the 31 transactions under original mechanism is 1,357,654. That is, the increment in gas consumption due to our mechanism is negligible (i.e., 0.01\%). Hence, the answer to RQ2 is:

\noindent
\fbox{	
	\parbox{0.95\textwidth}{%
		\emph{Our gas cost mechanism charges negligible additional gas from benign users.}
	}	
}

\subsection{Different Parameter Settings}
\label{sec_set}
We evaluate our mechanism under three different settings as listed in Table \ref{tab-set}. For example, ``3(5/1.2)'' means that in setting 3, $m$ and $\alpha$ are set to 5 and 1.2, respectively. Please note that the setting 2 is the default setting. Table \ref{tab-set} also presents the execution numbers of under-priced operations and the highest gas costs of them. For example, ``48/1,026,690'' in row 2, column 2 indicates \texttt{EXTCODECOPY} executes 48 times under setting 1 and the gas cost of the 48th \texttt{EXTCODECOPY} is 1,026,690. Please note that the gas cost of an operation keeps increasing if its execution number exceeds $\mu_i$ (Eqn. (\ref{eq_gas})).

\begin{table}[]
	\centering
	\scriptsize
	\vspace{-6ex}
	\caption{Execution numbers (before /) and the highest gas costs (after /) of under-priced operations under different settings}
	\vspace{-2ex}
	\label{tab-set}
	\begin{tabular}{|c|c|c|c|c|c|}
		\hline
		setting  & \texttt{BALANCE}      & \texttt{EXTCODECOPY}  & \texttt{EXTCODESIZE}  & \texttt{SLOAD}        & \texttt{SUICIDE}    \\ \hline
		1(1/5)   & 48/1,026,387 & 48/1,026,690 & 48/1,026,687 & 48/1,026,187 & 22/237     \\ \hline
		2(3/2)   & 99/456,819   & 99/457,122   & 99/457,119   & 99/456,619   & 81/37,640  \\ \hline
		3(5/1.2) & 329/135,590  & 328/131,052  & 328/131,049  & 329/135,390  & 289/31,440 \\ \hline
	\end{tabular}
\vspace{-6ex}
\end{table}

The experimental results demonstrate that our approach is sensitive to DoS attacks by setting a small $m$ and a large $\alpha$. The setting 1 detects attacks quicker (i.e., the execution numbers of under-priced operations are the lowest) than the other two settings. For example, the \texttt{EXTCODESIZE} attack executes 48 \texttt{EXTCODESIZE}, and its gas cost reaches 1,026,187 under the setting 1 whereas the attack executes 328 \texttt{EXTCODESIZE} and the gas cost of \texttt{EXTCODESIZE} reaches 131,049 under the setting 3. The results are as expected since the threshold $\mu_i$ depends on $m$ and $\alpha$ determines the speed of increasing gas costs.

\begin{figure}[ht]
	\centering
	\vspace{-4ex}
	\includegraphics[width=0.7\textwidth]{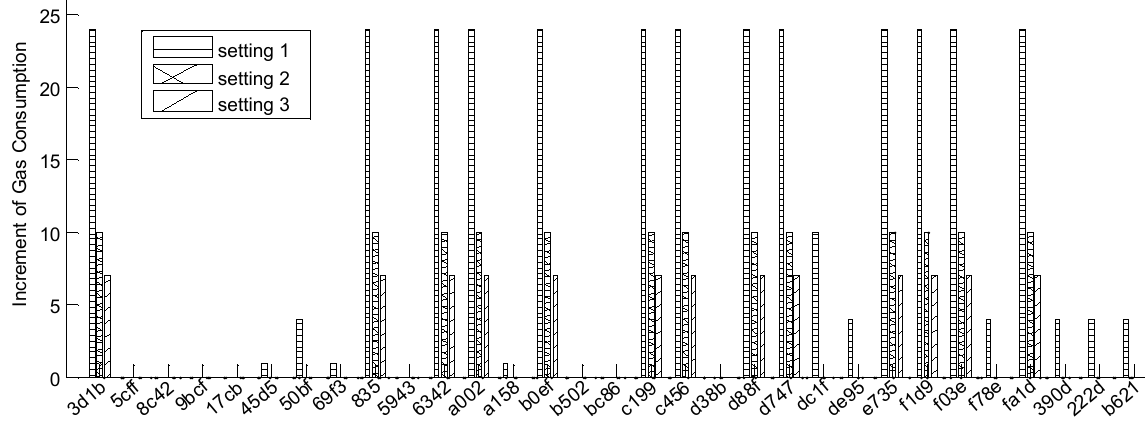}
	\vspace{-2ex}
	\caption{Additional gas consumption of 31 transactions under three different settings}
	\vspace{-4ex}
	\label{fig_increment}
\end{figure}

One may feel strange that \texttt{SUICIDE} presents different trend with the other attacks under different settings. For example, the gas cost of \texttt{SUICIDE} under the setting 2 is larger than that under the other two settings, whereas the gas costs of the other four under-priced operations under the setting 1 reach the largest value. The reason is that \texttt{SUICIDE} is not the most expensive operation during attack (as shown in Fig. \ref{fig_gas_attack}(b)), and thus the execution number of \texttt{SUICIDE} is influenced by the gas consumption of other expensive operations.
Fig. \ref{fig_increment} shows the increment in gas consumption of applying 31 transactions to the smart contract at 0x61F9d1cE56aC1623FeD4e949D7D420251fef0896 under three different settings. The $x$-axis specifies transactions in short, e.g., \emph{3d1b} is the first two bytes of a transaction hash which is 32 bytes in length. The results reveal that a setting that is more sensitive to DoS attacks may charge more execution fee from benign users.

We also evaluate whether our mechanism can defend against DoS attacks exploiting the five under-priced operations under the default setting with different window sizes. We compute $ave_i$ and $std_i$ of each operation $i$ for different windows sizes, including 100, 500, 1,000, 5,000 and 10,000.  We use the first 16,000 transactions since the lanuch of Ethereum for experiments, which do not include attacks. The attacks exploiting the five under-priced operations are conducted in our private chain for this experiment.

\begin{figure}[ht]
	\centering
	\vspace{-4ex}
	\includegraphics[width=0.6\textwidth]{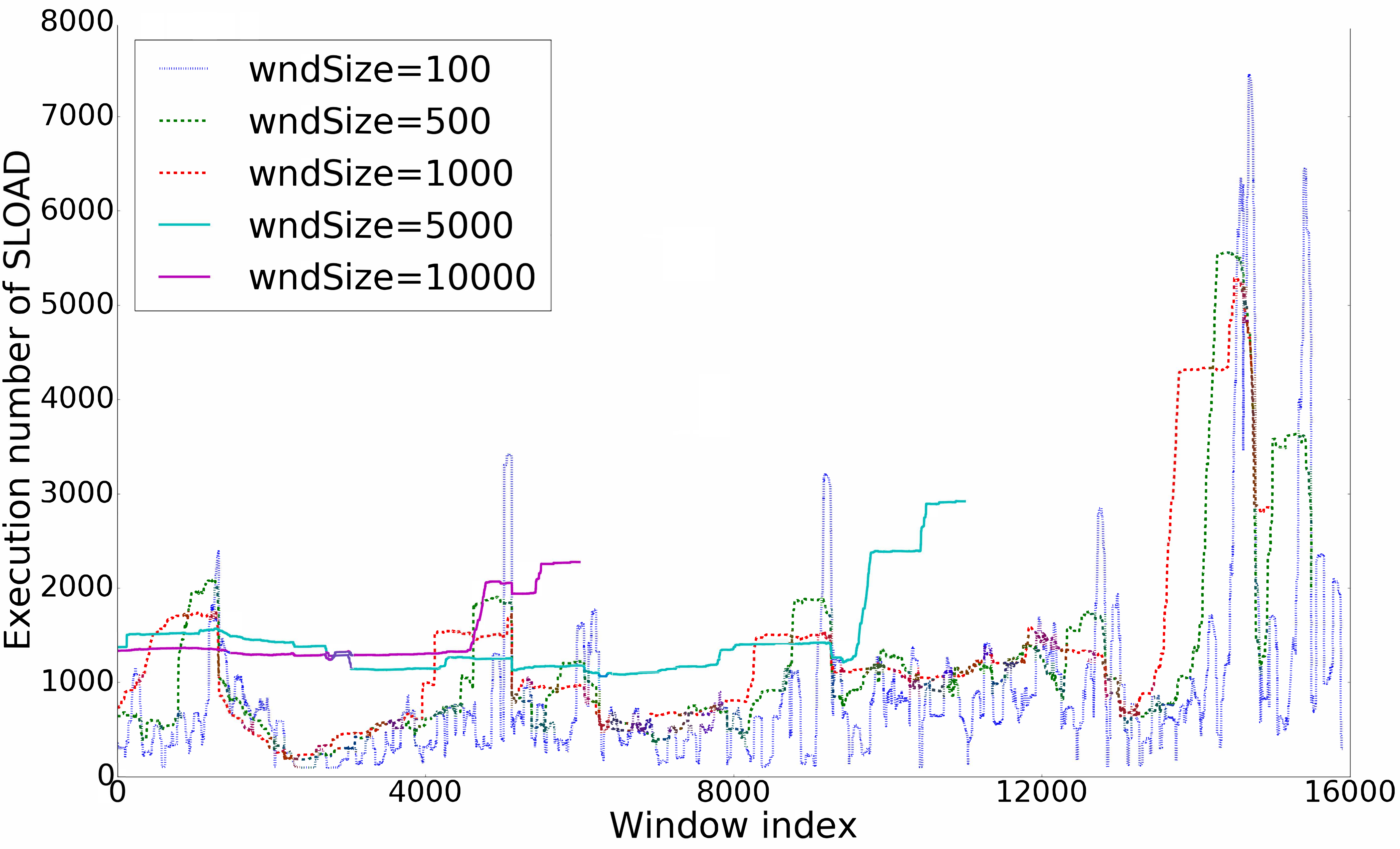}
	\vspace{-3ex}
	\caption{Execution numbers of \texttt{SLOAD} with different window sizes}
	\vspace{-5ex}
	\label{fig_sload}
\end{figure}

Fig .\ref{fig_sload} presents the execution numbers of \texttt{SLOAD} with different window sizes (the experiments of other four under-priced operations produce similar results). It shows that that our method is effective using the parameters computed from all window sizes because the under-priced operation executes more than the threshold $\mu_i$ at any window sizes and hence its gas cost increases during attacks. More precisely, the orginal gas cost method allows \texttt{SLOAD} to execute nearly 100,000 times (Fig. \ref{fig_number}) whereas our method reduces this number significantly. 

We assume all transaction in the windows for computing parameters are benign. Attackers may want to place crafted transactions into the windows to affect the process of computing $ave_i$ and $std_i$ for the sake of evading the detection. To make our approach more robust, we suggest analysts to set a relatively large window size (e.g., 10,000) that consists of many transactions. In another words, a large window size raises the difficulty for attackers to fill the window with crafted transactions and tamper parameters. Besides, after detecting an attacking transaction, we can filter out the attacking transactions in the windows by matching the transaction senders, attached data (specifying which function to call and providing augments) of transactions, the execution traces of contracts etc.

Hence, the answer to RQ3 is:

\noindent
\fbox{%	
	\parbox{0.99\textwidth}{%
		\emph{The experimental result show that DoS attacks can be detected quickly and negligible additional gas is introduced to benign users under different parameter settings. Our mechanism allows developers to easily adjust parameters with the evolving of Ethereum.}
	}%	
}
\section{Related Work}
\label{sec_related}
DoS attacks have posed a severe threat to the Internet\cite{Luo05,Xue14} and various systems\cite{Chen17,Jiang17} and services \cite{Tang14}. Although DoS attacks on Ethereum have been reported, there lacks of a systematic study on the attacks and the defense mechanisms. To the best of our knowledge, this paper presents the first work on defending against under-priced DoS attacks on Ethereum.

BLOCKBENCH~\cite{blockbench} is an evaluation framework for measuring the throughput, latency, scalability and fault-tolerance of private blockchains. Yasaweerasinghelage et al. propose to predict the latency of blockchain-based applications using architectural performance modeling and simulation tools~\cite{predict}. However, they~\cite{blockbench,predict} do not investigate the consumptions of computing resources for executing EVM operations.
OYENTE~\cite{oyente} is a symbolic executor for smart contracts which discovers four types of security vulnerabilities. GASPER~\cite{gasper}, based on OYENTE, finds under-optimized smart contracts automatically that cost more gas than necessary. A recent survey~\cite{survey} reports that smart contracts suffer from several kinds of vulnerabilities. One kind is \emph{gasless send}, indicating that a transaction sender may not consider the situation that sending Ether to another account is possible to fail due to the out-of-gas exception. Sergey et al. reveal that smart contracts will suffer from similar problems that often occur in transitional concurrent programs~\cite{concurrency}. However, they~\cite{oyente,gasper,survey,concurrency} do not consider DoS attacks to Ethereum, which exploit under-priced EVM operations.

Verification is used for verifying the runtime safety and functional correctness of smart contracts. Bhargavan et al. propose to translate a smart contract into F$^\star$, a functional language before formal analysis~\cite{formal}. Similarly, Pettersson and Edstr\"{o}m suggest developing smart contracts in Idris, a functional language, and using type system to capture errors at compile time~\cite{idris}. Hirai formally defines EVM in Lem, an intermediate language similar to a functional language, facilitating further analysis and generation of smart contracts~\cite{evm}. However,  they neither verify nor detect DoS attacks due to under-priced operations~\cite{formal,idris,evm}.
Hawk is a smart contract system protecting transactional privacy~\cite{hawk}. Town Crier~\cite{town} aims at providing trustworthy data to smart contracts since they need data out from the blockchain. Juels et al. report that smart contracts can be used to commit crimes, such as privacy leakage, theft of cryptographic keys~\cite{ring}. However, they~\cite{hawk,town,ring} do not discuss the threats resulting from Ethereum DoS attacks.

\section{Conclusion}
\label{sec_conclusion}
We investigate the gas cost setting in Ethereum because it could be exploited to launch DoS attacks. By proposing an emulation-based framework to automatically measure the resource consumptions of EVM operations, we find that Ethereum does not assign proper gas costs to operations and it is difficult to properly assign fixed gas costs to operations for defending against known and unknown DoS attacks. Therefore, we propose a DoS-resistant gas cost mechanism, which dynamically adjusts the costs of operations according to the number of executions. Our approach is flexible and secure, and we design a special smart contract that collaborates with the customized EVM to avoid frequently updating Ethereum client. Experimental results show that our method effectively thwarts known and unknown DoS attacks, and introduces negligible additional gas to benign users. 

\section*{Acknowledgment}

This work was supported in part by the National Natural Science Foundation of China, No.61402080, No.61572115, No.61502086, No.61572109, China Postdoctoral Science Foundation founded project, No.2014M562307, and Shenzhen City Science and Technology R\&D Fund (No. JCYJ20150630115257892).

\bibliographystyle{splncs_srt}
\bibliography{ref}

\end{document}